\documentclass[12pt,preprint]{aastex}
\shorttitle{TRIPLE STAR STABILITY}
\shortauthors{Khodykin, Zakharov, \& Andersen}
\begin{document}

\title{STABILITY OF TRIPLE STAR SYSTEMS WITH HIGHLY INCLINED ORBITS}
\author{S. A. Khodykin}
\affil{Volgograd Pedagogical University}
\affil{Volgograd, Russia}
\email{khodykin@avtlg.ru}
\author{A. I. Zakharov}
\affil{Sternberg Astronomical Institute}
\affil{Moscow University, Moscow 119899, Russia}
\email{zakh@sai.msu.ru}
\author{W. L. Andersen}
\affil{Eastern New Mexico University}
\affil{Portales, N.M.}
\email{William.Andersen@enmu.edu}

%\slugcomment{Submitted to the Astrophysical Journal on 16 February 2003}

\begin{abstract}

It is well established that certain detached eclipsing binary stars 
exhibit apsidal motions whose values are in disagreement with 
calculated deviations from Keplerian motion based on tidal effects 
and the general theory of relativity. Although many theoretical 
scenarios have been demonstrated to bring calculations into line 
with observations, all have seemed unlikely for various reasons.
In particular, it 
has been established that the hypothesis of a third star 
in an orbit almost perpendicular to the orbital plane of the close 
binary system can explain the anomalous motion at least in some cases. 
The stability of triple star systems with highly inclined orbits
has been in doubt, however.

We have found conditions that allow the
long-term stability of such systems so that the third-body hypothesis
now seems a likely resolution of the apsidal motion problem.
We apply our stability criteria to the cases of AS Cam and DI Her
and recommend observations at the new Keck interferometer which
should be able to directly observe the third bodies in these systems.

\end{abstract}




\section{Introduction}

Discrepancy between observation and theoretical predictions of
the apsidal motion of certain detached binary stars has remained
an outstanding problem for two decades. In the cases of
AS Cam and DI Her, for example, observed apsidal motion rates
are a fraction of the theoretical predictions based on
stellar structure, tidal, and relativistic effects.

As was first pointed out by \citet{rud59}, the effect of relativistic 
gravity is significant in the case of a number of detached binary stars. 
Although he considered only DI~Her in detail, the number of 
interesting cases has grown to about half a dozen 
\citep{koc77,mof84}, including AS Cam \citep{mal89}.

The discovery of anomalous apsidal motion by \citet{mar80} was 
initially considered a possible challenge to general relativity. 
\citet{mof84} invented an alternative theory of gravity that 
harbored 
differing predictions for the apsidal motion of binary stars and yet
maintained agreement with other tests of general relativity. The 
predictive power of this
theory is weakened by the existence of a new adjustable parameter for 
each star. Moreover, increasingly severe tests of general 
relativity such as in \citet{tay89} make such large deviations at AU scales seem 
unlikely.
Several other, less exotic, solutions have been 
proposed. In one scenario the rapid circularization of the orbit 
occurs because of dissipation of angular momentum from stellar 
oscillations or a large amount of mass loss. Another reasonable guess 
is that the close binary system (CBS) orbit is surrounded by a resisting medium in the form 
of gas clouds. The required density well exceeds observational limits 
in the case of DI~Her, however. These and other alternatives are reviewed
in \citet{gui85}, \citet{mal89}, and
\cite{cla97,cla98}. It is the purpose of this paper to demonstrate the 
feasibility of one particular 
explanation that has been put forward in the literature. Our 
considerations may also find application in other triple star or 
suspected triple star systems such as in \citet{coe02} where a triple 
star model for the X-ray pulsar AX J0051-733 is proposed to explain 
puzzling features of the spectrum of a candidate optical counterpart.

It is well known that the hypothesis of third stars in outer orbits
of these close binary systems can bring theory into line with observed
apsidal periods \citep{kha91,kho97}, but the stability of such triple star systems has
been in doubt \citep{har68}. We show here that the inclusion of the 
apsidal motion as an additional perturbation leads to the conclusion 
that such triple star motions can be stable. In particular, the 
triple star models of \citet{kho97} and \citet{kha91}
which reconcile the cases 
of AS Cam and DI Her with observations are shown to be stable.
Furthermore, we show that 
observations utilizing the new Keck interferometer should be able to 
directly image the putative third bodies in these two systems.

\section{The Lagrange Planetary Equations for the Hierarchical Three 
Body System}

\begin{figure}
\plotone{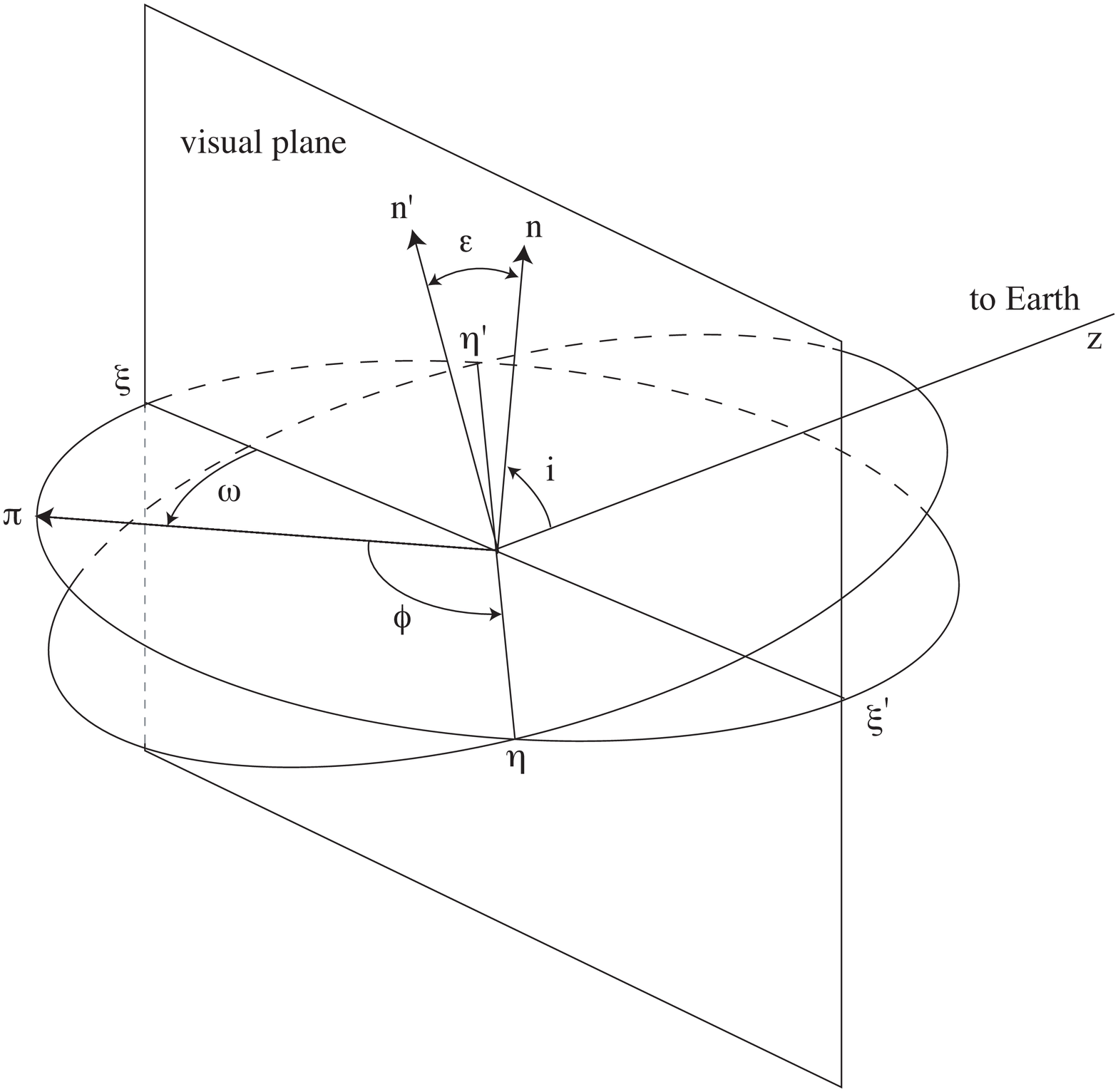}
\caption{Kinematic variables describing the 
relative orientation of the orbits of the CBS and the third 
body. \label{f1}}
\end{figure}

We have studied numerically the dynamical evolution of a hierarchical 
triple system consisting of a massive CBS and a third star 
of moderate mass. Figure~\ref{f1} shows the notation used.
The calculations were done perturbatively using the disturbing 
function method \citep{kop78}. We have assumed that the three stars are pointlike and 
isolated from other stars. We ignore internal dynamical exchanges 
such as synchronization, angular momentum 
exchange, and orbital precession of the CBS.
Classical tidal effects and relativistic effects are assumed to be independent 
and additive.

The disturbing functions for 
the CBS and the third-body are adapted from  
\citet[p. 14]{bro33}. They are, respectively,

\begin{eqnarray}
R&=&4\pi^{2}{{m_{3}}\over{r}}\left({r\over 
r'}\right)^{3}\sum_{n=1}^{\infty}{{m_{1}^{n}-(-m_{2})^{n}}\over{(m_{1}+m_{2})^{n}}}\left({r\over r'}\right)^{n-1}P_{n+1}\\ 
R^{tb}&=&{{{m_{1}+m_{2}+m_{3}}\over{(m_{1}+m_{2})^{2}}}{{m_{1}m_{2}}\over{m_{3}}}R}
\end{eqnarray}

We average $R$ and $R_{tb}$ over the mean anomalies of the CBS and 
third-body (hence, twice averaged).
The first-order terms of the twice-averaged disturbing functions, $R_2$ and $R^{tb}_{2}$, are 

\begin{eqnarray}
R_{2}&=&{\pi^{2}q' 
m_{1}a^{2}\over{2(1-e'^{2})^{3/2}a'^{3}}}[3N^{2}(1-e^{2})-15e^{2}Q^{2}+(6e^{2}-1)]\\ 
R^{tb}_{2}&=&{{q(1+q+q')\over{q'(1+q)^{2}}}R_{2}}
\end{eqnarray}

where the masses are included via the ratios $q=m_{2}/m_{1}$ and 
$q'=m'/m_{1}$. We choose our units of measurement to be AU, years, and 
solar masses, so that the Newtonian gravitational 
constant is $G=4\pi^{2}$. The orientation of the third-body orbital plane with respect 
to the close binary orbital plane is described by the direction 
cosines ($Q$,$S$,$N$) of the unit vector normal to the third-body orbital 
plane. We refer the direction cosines to the periastron, a 
perpendicular to the periastron, and the direction normal to the close 
binary orbit. Let $\epsilon$ be the angle between the two orbital 
planes, 
and call the angle measured from the periastron to the line of 
intersection of the orbital planes $\phi$. We then have

\begin{eqnarray}
Q&=&\sin\epsilon \sin\phi \label{eq:Q} \\ 
S&=&-\sin\epsilon \cos\phi \label{eq:S} \\
N&=&\cos\epsilon \label{eq:N}
\end{eqnarray}

Since we are averaging over the mean anomaly, there is no Lagrange 
planetary equation for $M$ and furthermore the semimajor axis $a$ has 
no time dependence. The planetary equations for the perturbation of 
the remaining CBS orbital elements by the third-body are

\begin{eqnarray}
    {({{da}\over{dt}})}_{tb}&=&0 \label{eq:aMotion} \\
{({{de}\over{dt}})}_{tb}&=&5AeQS = -{5\over2}A e \sin^{2}\epsilon \sin 
2\phi \label{eq:eMotion} \\
{({{d\omega}\over{dt}})}_{tb}&=&A\left[2-5Q^{2}-N^{2}-N\cot i 
\left({{1+4e^{2}\over{1-e^{2}}}Q\sin\omega}+S \cos 
\omega\right)\right] 
\label{eq:omegaMotion} \\
{({{di}\over{dt}})}_{tb}&=&AN\left[{{1+4e^{2}\over{1-e^{2}}} Q \cos\omega}-S 
\sin\omega\right] \label{eq:iMotion} \\
{({{d\Omega}\over{dt}})}_{tb}&=&A N \csc i  \left[ S \cos\omega
+{{1+4e^{2}}\over{1-e^{2}}} Q \sin\omega) \right]  \label{eq:OmegaMotion}
\end{eqnarray}
where
\begin{eqnarray}
    A&=&{3\pi (1-e^{2})^{1\over 2}P q'}\over{2(1-e'^{2})^{3\over 
    2}P'^{2}(1+q+q')}
\end{eqnarray}

The planetary equations for the perturbation of 
the third-body orbital elements are

\begin{eqnarray}
    {({{da'}\over{dt}})}_{tb}&=&0 \label{eq:aPrimeMotion} \\
{({{de'}\over{dt}})}_{tb}&=&0 \label{eq:ePrimeMotion} \\
{({{d\omega'}\over{dt}})}_{tb}&=&B\left[{{3e^{2}+2}\over{2(1-e^{2})}}-
{3\over2}\left({{1+4e^{2}\over{1-e^{2}}}}Q^{2}+S^{2}\right)+
\cot i' \left({{1+4e^{2}\over{1-e^{2}}}Q T}+S U\right)\right] 
\label{eq:omegaPrimeMotion} \\
{({{di'}\over{dt}})}_{tb}&=&-B \left[{{1+4e^{2}\over{1-e^{2}}} Q F}+S G\right] \label{eq:iPrimeMotion} \\
{({{d\Omega'}\over{dt}})}_{tb}&=&-B\csc i' \left[{{1+4e^{2}\over{1-e^{2}}} 
Q T}+S U\right]  
\label{eq:OmegaPrimeMotion}
\end{eqnarray}
where
\begin{eqnarray}
    B&=&{3\pi q (1-e^{2})P^{4\over 
    3}}\over{2(1-e'^{2})^{2}P'^{7/3}(1+q)^{4/3}(1+q+q')^{2/3}}
\end{eqnarray}

In writing equations~(\ref{eq:omegaPrimeMotion}-- 
\ref{eq:OmegaPrimeMotion}) we have introduced the direction cosines 
of the nodal line of the third-body referred to the same CBS axes by
which $Q$,$S$, and $N$ are defined 
above. These cosines are

\begin{eqnarray}
F&=&\sin(\Omega'-\Omega)\cos i \sin\omega + \cos(\Omega'-\Omega) 
\cos\omega \label{eq:F} \\ 
G&=&\sin(\Omega'-\Omega)\cos i \cos\omega - \cos(\Omega'-\Omega) 
\sin\omega \label{eq:G} \\
H&=&\sin(\Omega'-\Omega)\sin i \label{eq:H}
\end{eqnarray}

Finally, we use cosines of the direction perpendicular to the third 
body nodal line and behind the visual plane. These cosines are

\begin{eqnarray}
T&=&\left[\cos(\Omega'-\Omega)\cos i \cos i' + \sin i \sin 
i'\right]\sin \omega - \sin(\Omega'-\Omega)\cos i' \cos\omega \label{eq:T} \\ 
U&=&\left[\cos(\Omega'-\Omega)\cos i \cos i' + \sin i \sin 
i'\right]\cos \omega + \sin(\Omega'-\Omega)\cos i' \sin\omega \label{eq:U} \\
V&=&\cos(\Omega'-\Omega)\sin i \cos i' - \cos i \sin i' \label{eq:V}
\end{eqnarray}

\section{An Alternative Formulation of the Equations of Motion}

It is worth noting that the equations of motion can be written in a 
particularly elegant fashion by noting that they determine nothing 
more than an 
instantaneous rotation that can be referred to the CBS coordinates.
%which we shall call $X$, $Y$, and $Z$.
The problem is then to 
express $d\omega/dt$, $di/dt$, $d\phi/dt$, and $d\epsilon/dt$ in terms 
of this angular velocity vector\citep[p. 174]{gol02}. Referring to this angular velocity as 
$\mathbf{\Psi}$, we first resolve it into three components: 
$\mathbf{\Psi}=\mathbf{\Psi}_{\omega}+\mathbf{\Psi}_{i}+\mathbf{\Psi}_{z}$.
These components are related by rotation(s) to the angular 
velocity components.
For example,
$\mathbf{\Psi}_{i}$ is just $-di/dt$ referred to the nodal line $\xi\xi'$.
Referring to figure~\ref{f1}, we see that a rotation by $\omega$ about the $z$-axis brings
this angular velocity into the CBS 
system; thus

\begin{eqnarray}
    \left(
    \begin{array}{c} X \\ Y \\ Z  \end{array}
    \right)
    =
    \left(
    \begin{array}{ccc}
	\cos\omega & \sin\omega & 0 \\
	-\sin\omega & \cos\omega & 0 \\
	0           &            & 1
    \end{array}
    \right)
    \left(
	\begin{array}{c} -{{di}\over{dt}} \\ 0 \\ 0  \end{array}
    \right)
\end{eqnarray}

where $X$, $Y$, and $Z$ are the components of $\mathbf{\Psi}$ in the 
CBS coordinates.
Proceeding in this manner, one obtains three equations giving $X$, $Y$, 
and $Z$ in terms of $d\omega/dt$, 
$di/dt$, and $d\Omega/dt$. These 
equations are inverted to yield

\begin{eqnarray}
{{d\omega}\over{dt}}&=& \cot i [ X \sin\omega + Y \cos\omega ] + Z 
\label{eq:omegaMotionAlter} \\
{{di}\over{dt}}&=& -X\cos\omega +Y\sin\omega \label{eq:iMotionAlter} \\
{{d\Omega}\over{dt}}&=& -\csc i [ X\sin\omega + Y\cos\omega ]  
\label{eq:OmegaMotionAlter}
\end{eqnarray}

Similar considerations for a circular third-body orbit result 
in

\begin{eqnarray}
{{d\epsilon}\over{dt}}&=& X\cos\phi +Y\sin\phi \label{eq:epsilonMotionAlter} \\
{{d\phi}\over{dt}}&=& -\cot\epsilon ( X\sin\phi - Y\cos\phi ) + Z.
\label{eq:phiMotionAlter}
\end{eqnarray}

Comparing 
equations~(\ref{eq:omegaMotionAlter}--\ref{eq:OmegaMotionAlter})
with (\ref{eq:omegaMotion}--\ref{eq:OmegaMotion}), we determine the rate of
rotation of the CBS frame, as measured by CBS 
coordinates, to be

\begin{eqnarray}
    (X,Y,Z)=-A\left( N Q {{1+4e^{2}}\over{1-e^{2}}} , N S, 
    N^{2}+5Q^{2}-2 \right)
    \label{eq:rotRate}
\end{eqnarray}

%\begin{eqnarray}
%    \begin{matrix} $\mathbf{\Psi})_{X}$ & $\mathbf{\Psi})_{Y}$ & 
%    $\mathbf{\Psi})_{Z}$ \end{matrix} = 
%    \begin{matrix}
%	\cos\omega & \sin\omega & 0 \\
%	-\sin\omega & \cos\omega & 0 \\
%	0           &            & 1 
%    \end{matrix}
%    \begin{array}{c} (-{{di}\over{dt}} \\ 0 \\ 0 )\end{array}
%\end{eqnarray}

It has been pointed out by \citet{kis98} (using different angles 
to define orbit orientations) that this formulation admits 
two exact integrals and leads to a first-order elliptical differential 
equation for the eccentricity. Our calculations, however, are 
entirely numerical and are based on equations
(\ref{eq:aMotion}--\ref{eq:OmegaMotion}) and
(\ref{eq:aPrimeMotion}--\ref{eq:OmegaPrimeMotion}).

\section{The Instability Problem of Hierarchical Triple Star Systems}

\begin{figure}
\plotone{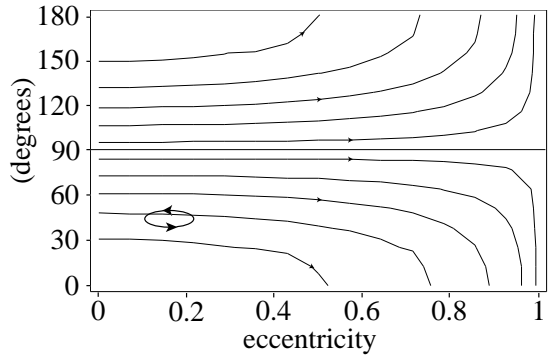}
\caption{Evolution of $\epsilon$ and $e$ due to 
angular momentum exchange between the CBS and the third-body. The 
loop schematically illustrates the results of numerical integration of the 
equations of motion of AS Cam including the influence of the CBS apsidal motion 
as an additional perturbation.\label{f2}}
\end{figure}

There are a couple of ways of seeing the problem of instability. One 
approach utilizes the conservation of angular momentum. We assume that
$i\approx \pi/2$, so that the $\cot i$ term in equation~(\ref{eq:omegaMotion})
can be ignored. This leads to a
large eccentricity excursion due to the angular momentum exchange between 
the CBS and the third-body. Since the disturbing function $R^{tb}_{2}$ 
depends on neither $M'$ (the mean anomaly) nor $\omega'$ (the 
longitude of the periastron of the third-body with respect to the 
ascending node),
there is no secular variation 
of the third-body semimajor axis $a'$ and eccentricity $e'$. 
Therefore, the orbit of the third-body maintains its shape, and the 
magnitude of its orbital angular momentum $\bf L'$ is a constant of motion. 
The direction of the third-body orbital angular momentum, however, will 
change, as 
the following argument shows. The orbital angular momentum of the 
close binary system ${\bf L_{BS}}$ is 3--10 times smaller 
than ${\bf L'}$ for the systems under investigation. Since the 
rotational angular momentum of the stars is about 2 orders of 
magnitude less than ${\bf L'}$, the total angular momentum of the 
system is ${\bf L_{TOT}=L_{BS}+L'}$.
Calculations reveal that the CBS angular momentum is transferred to 
the third star. Since the magnitude of $\bf L'$ can not change, this 
angular momentum transfer forces a change in the orientation of the 
third-body orbit with respect to the total angular momentum.
Conservation of angular momentum dictates the connection between
$\epsilon$ and $e$ shown in figure~\ref{f2}. As angular momentum is 
transferred, the coordinate values $(e ,\epsilon)$ slide along one 
of the integral curves determined from the initial values of the 
eccentricities $e$ and $e'$, mass ratios $q$ and $q'$, relative 
inclination $\epsilon$ and ratio of the semimajor axes $a/ 
a'$. In the presence of a small but nonzero $\cot i$ the above 
argument does not take the orbit all the way to $e=0$, but the 
eccentricity may become close enough to unity that tidal effects or even 
collision destroy the system. This turns out to be the case for 
DI~Her and AS~Cam.

We can gain additional insight into the dynamics of the eccentricity by looking directly
at the equations of motion. More specifically, we look at equation
(\ref{eq:omegaMotion}) neglecting the $\cot i$ term and further assume 
that the motion of the nodal line $\eta\eta'$ caused by nutation and 
precession of the orbits is small. Under these assumptions,

\begin{eqnarray}
{({{d\omega}\over{dt}})}_{tb}&\approx&A\left[2-5Q^{2}-N^{2}\right] \label{eq:omegaMotionNoCoti} \\
{({{d\omega}\over{dt}})}_{tb}&\approx&-{({{d\phi}\over{dt}})}_{tb} 
\label{eq:phiMotion}
\end{eqnarray}

First, we note that for $\epsilon\le 
30\degr$ we have $Q, S<<1$.
We see then that the righthand side of equation~(\ref{eq:omegaMotion})
is always positive. Thus $\omega$ and $\phi$ change 
monotonically and run over all quadrants, never coming to rest. Therefore, the perturbations 
in the eccentricity (equation~(\ref{eq:eMotion})) are periodic, and the 
eccentricity suffers no excursion to a value nearing unity.
The situation is much different if $\epsilon > 30 \degr$. At such 
inclinations, equation~(\ref{eq:omegaMotion}) implies in general four 
values of $\phi$ for which $d\omega/dt=0$. Let us examine the motion 
of $\phi$ at these roots.
Equations (\ref{eq:omegaMotionNoCoti}) and (\ref{eq:phiMotion}) 
together imply

\begin{eqnarray}
{({{d^{2}\phi}\over{dt^{2}}})}_{tb}&\sim&\sin^{2}\epsilon \sin 2\phi {({{d\phi}\over{dt}})}_{tb} 
\label{eq:phiAcceleration}
\end{eqnarray}

This tells us that the four points of stationary $\omega$ (and $\phi$) 
will be stable only if $\sin 2\phi < 0$. Looking at equation 
(\ref{eq:eMotion}) we see that this same condition guarantees
$(de/dt)_{tb}>0$. Thus $\phi$ will wander until it reaches a value 
that results in an eccentricity excursion. Again, we emphasize that 
these semi-quantitative arguments are limited to the extent that we 
neglect the $\cot i$ term in equation (\ref{eq:omegaMotion}).

%Closer inspection reveals 
%that two of these roots represent unstable equilibria and two roots 
%are stable. Thus, the orbit of the CBS rotates 
%about the direction $\bf n$ until it reaches a point of stable 
%equilibrium, whereupon it stops. For these stable values of $\phi$, 
%equation~(\ref{eq:eMotion}) implies $de/dt>0$. Thus, the eccentricity 
%grows to a value which may destroy the CBS by virtue of
%collision or tidal interactions.

The characteristic time scale for
the change of eccentricity can be obtained from 
equation~(\ref{eq:eMotion}):

\begin{eqnarray}
    \tau_{e}&\approx&{{1-e}\over{\big({{de}\over{dt}}\big)}}\approx 0.1 
    \sqrt{{1-e}\over{1+e}}{{(1-e'^{2})^{3/2}a'^{3}}\over{e P m'}} 
    \label{eq:etime}
\end{eqnarray}

For DI Her and AS Cam these times are about 700 and 400 yr 
respectively, as was confirmed directly by numerical integration. Thus 
at first glance the conclusion seems to be that only nearly coplanar hierarchical 
triple systems can be stable for more than a few hundred years. If 
this conclusion were correct, the third-body hypothesis would be
eliminated as a probable solution 
to the apsidal motion discrepancy.

\section{A Possible Resolution of the Problem of Instability}

We now explore the consequences of including the effect of
stellar structure (namely tidal-rotational deformation of the CBS pair)
and the relativistic effect as additive perturbations on the motion 
of $\omega$. We assume that the structure effect $(d\omega /dt)_{cl}$ and the 
relativistic effect $(d\omega /dt)_{rel}$ act in simple superposition 
with the 
effect of the third-body $(d\omega /dt)_{tb}$, so 
that their influence can be represented by simply adding them to 
$(d\omega/dt)_{tb}$. If the combined effect of 
these two additional terms is 
of the same order or greater than the third-body effect, that is, if

\begin{eqnarray}
    {\left({{d\omega}\over{dt}}\right)}_{tb}&\stackrel{\textstyle
    <}{\sim}&{\left({{d\omega}\over{dt}}\right)}_{cl}+{\left({{d\omega}\over{dt}}\right)}_{rel}
    \label{eq:omegacondition}
\end{eqnarray}

then the motion of $\omega$ will not stop. Thus $\omega$ and $\phi$ 
will change monotonically, resulting in periodic perturbations of the 
orbital elements of the CBS leading to stability, as in the case of 
low inclinations discussed above.
 
We can derive stability criteria on the basis of 
equation~(\ref{eq:omegacondition}).
We consider the cases in which either the 
classical deformation effect or the relativistic effect dominates 
using well-known relationships for $(d\omega /dt)_{cl}$ from \citet{kop78}
and $(d\omega /dt)_{rel}$ from \citet{rud59} and \citet{mar80}.

If the classical effect dominates in the sense that 
$|(d\omega/dt)_{tb}|<(d\omega/dt)_{cl}$, we have

\begin{eqnarray}
S^{*}_{tb}&=&{(1-e^{2})^{7}\over (1-e'^{2})^{3/2}}\left({a\over a'}\right)^{3}
{q' \over q(1+q)}\left({3\over 4} - \cos^{2}\epsilon \right) 
<S_{cl}=10k_{2}r^{5} \label{eq:classdom}
\end{eqnarray}

If the relativistic effect is predominant 
[$|(d\omega/dt)_{tb}|<(d\omega/dt)_{rel}$], then

\begin{eqnarray}
S^{**}_{tb}&=&{(1-e^{2})^{3/2}\over (1-e'^{2})^{3/2}}\left({a\over a'}\right)^{3}
{q'a\over M_{1}(1+q)^{2}}\left({3\over 4} - \cos^{2}\epsilon \right)
<S_{rel}={G\over c^{2}}=10^{-8} \label{eq:reldom}
\end{eqnarray}

Table~\ref{Table1} displays the application of these criteria to AS Cam 
and DI Her. We see that, according to our hypothesis, both of these 
supposed triple star systems are predicted to be stable.
The stability of AS Cam is provided by the 
classical effect whereas DI Her is stable because of the relativistic 
apsidal motion.

\begin{figure}
\plottwo{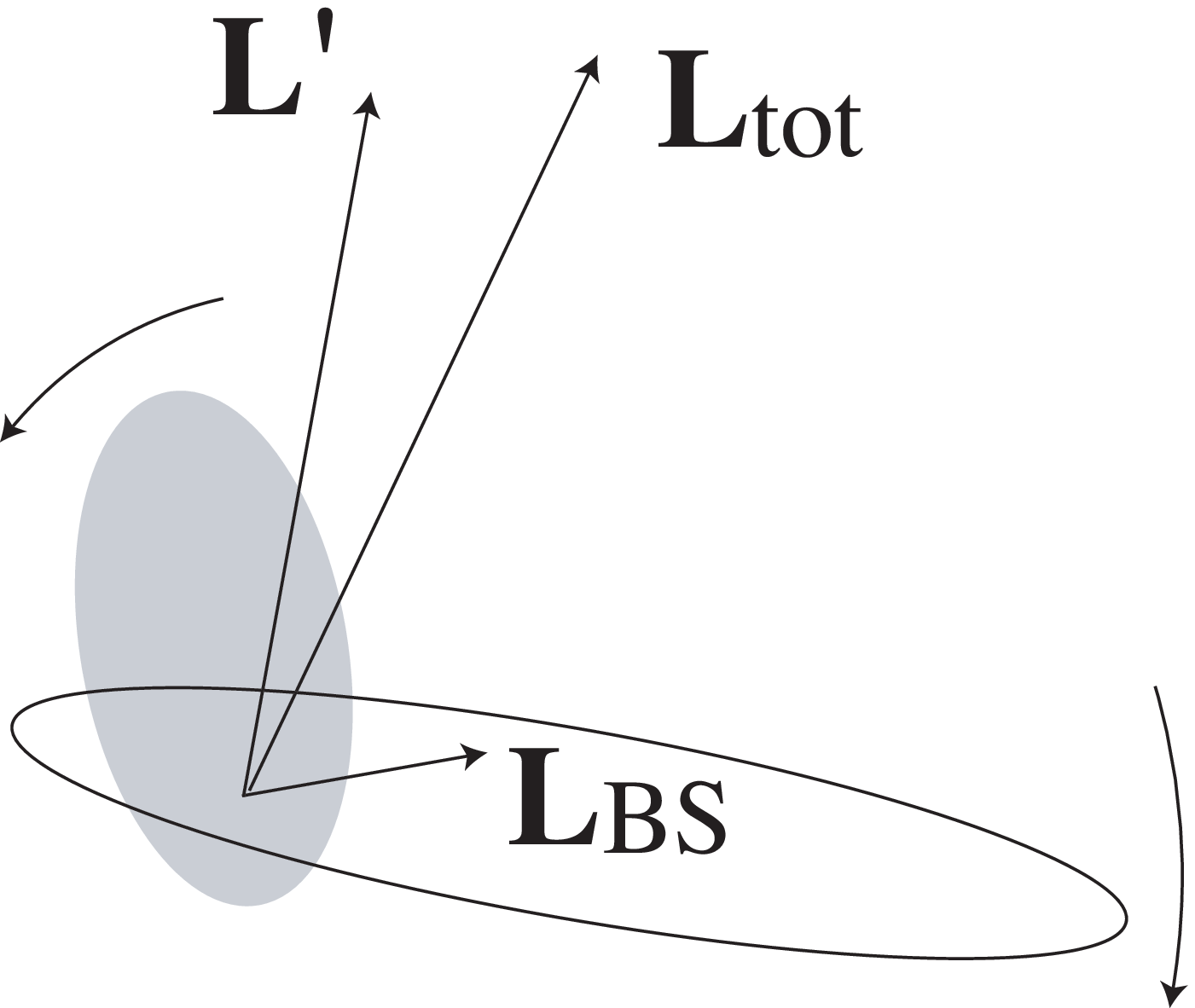}{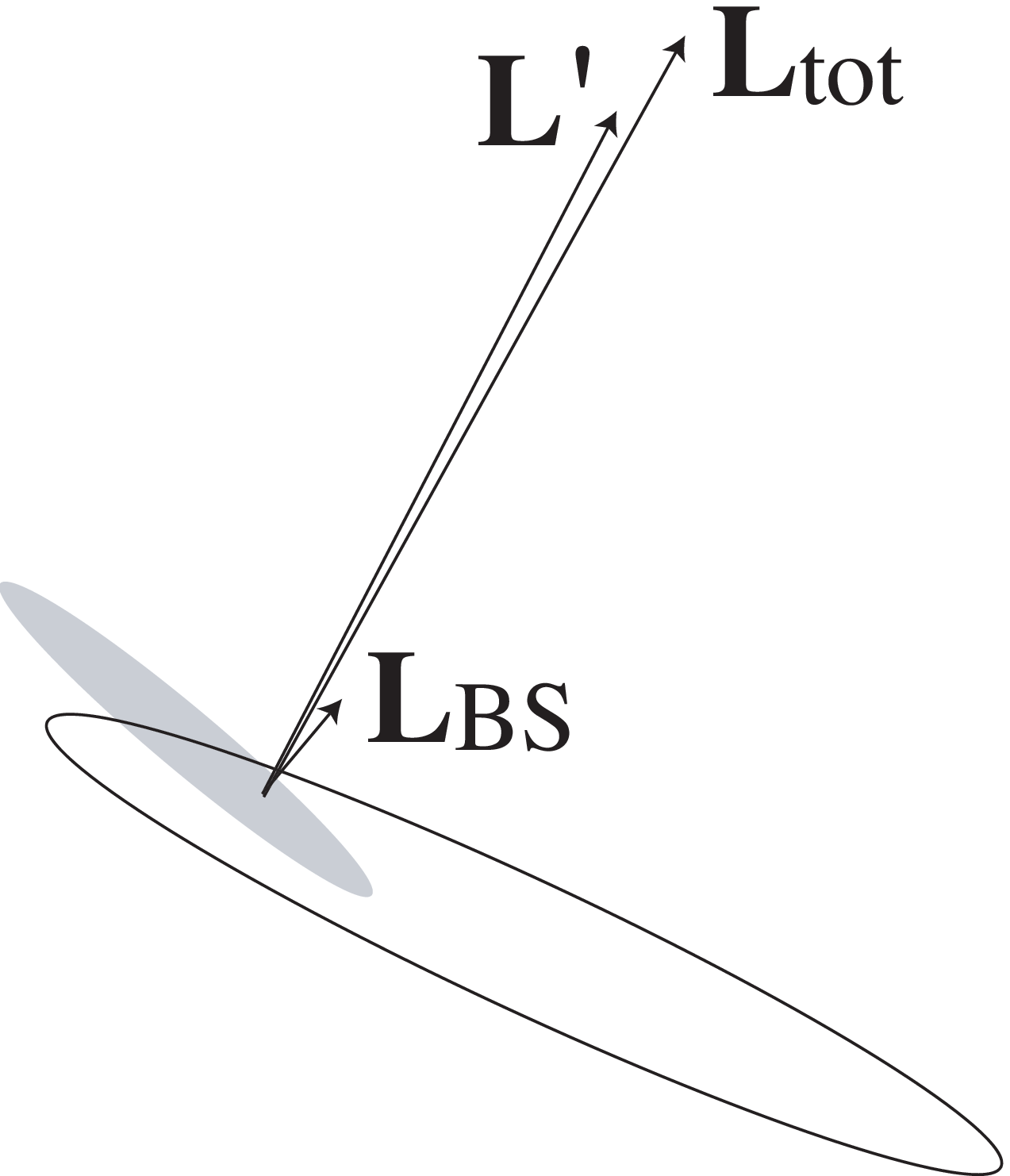}
\caption{Motion schematic. As the CBS loses angular momentum, its 
orbit overturns and eccentricity increases to a value close to unity.
If the apsidal motion 
of the CBS is included as an additional perturbation, then $\epsilon$ 
and $e$ oscillations are moderated, and the triple system orbits move back and forth
like a butterfly.\label{f3}}
\end{figure}

We have confirmed this behavior by means of numerical integrations of 
the equations of motion.
The calculations were done perturbatively using the disturbing 
function method \citep{kop78}. We have assumed that the stars are pointlike, and 
the close encounters are ignored. We ignore internal dynamical exchanges 
such as synchronization, angular momentum 
exchange, and orbital precession of the CBS. The apsidal motion in the CBS caused by
both the classical tidal-rotational 
deformation of the components and the relativistic apsidal motion is 
described by disturbing functions as in \citet{kha91}.
The classical tidal effects and relativistic effects are assumed to be independent 
and additive. The results for AS Cam are shown in 
figure~\ref{f2}. 
Thus the angular momentum is transferred back and forth between the 
third-body and the CBS. These periodical variations in $e$ and 
$\epsilon$ can be visualized as a flapping of the CBS and third-body 
orbits, almost like a butterfly (figure~\ref{f3}).

\section{Discussion}

We begin our discussion by contrasting our stability criteria with those of 
\citet{roy79}, \citet{sze77}, and \citet{egg95} for the case
of DI~Her using the same orbital parameters as \citet{kha91}.

\citet{roy79} allows a very close orbit of the third-body, the 
restriction being only that the semimajor axis satisfy $a'\geq 
0.3~\rm{AU}$ corresponding to a period $P' \geq 18.3~{\rm days}$.
This seems much too close to the inner binary for stability at any 
inclination of the third-body orbit.

Although \citet{sze77} deal primarily with the case of coplanar orbits, they 
indicate how their results may be extended to third-body 
orbits inclined to the inner binary orbit. We have applied their 
stability criteria assuming that the third-body orbit is perpendicular 
to the inner binary orbital plane. The result is that stability 
criterion requires $a'\geq 2.9~\rm{AU}$ ($P' \geq 1.5~{\rm yr}$).
This seems more reasonable, but in fact numerical simulations suggest 
that this is still too close \citep{kho97}.

Finally, \citet{egg95} predict that the system is stable for any 
third-body orbit with $a'\geq 1.2~\rm{AU}$ ($P' \geq 0.39~{\rm yr}$).

The criteria we propose prove more restrictive. From 
equation~(\ref{eq:reldom}) we determine that $a'\geq 9.5~\rm{AU}$
($P' \geq 9~{\rm yr}$). We have confirmed the stability of the 
purported three body system of DI~Her under this restriction on the 
third-body orbit. We also wish to emphasize that the criteria of
\citet{roy79}, \citet{sze77}, and \citet{egg95} are all based on the
purely Newtonian gravitational theory of point mass orbits. Our 
criteria depend on structure effects and/or general relativity. In the 
case of DI~Her the effect of general relativity dominates. It appears
that the stability of the hypothetical three body system of DI~Her is 
to be found in the physics of general relativity rather than 
structure effects. Thus general relativity itself provides 
stability to the three-body model of DI~Her, which seems necessary to 
bring its theoretical apsidal motion into line with observations. A 
similar stabilizing role for general relativity was found
by \citet{hol97} in their investigations of 
a planet orbiting one star of a binary system. Our criteria are 
somewhat more general in that structure effects are also considered. 

We close with a discussion of the prospects of making direct 
observations of these hypothetical third-body companions of AS~Cam 
and DI~Her in light of recent advances in optical interferometry and adaptive 
optics.

Indirect evidence for a third-body in AS Cam (B+B9.5, P=3.43\,days, 
e=0.17, V=8.6) has already been
found by \citet{koz99a} and \citet{koz99b}, who found, imposed upon the timing of eclipse 
minima, a cyclic variation with a period of 2.2 yr. They have 
interpreted this signal as due to the Roemer-like influence of a 
third star. The calculations of 
\citet{kho97} indicate that in order to account for the anomalous
apsidal motion of AS Cam, the third-body should be of about $1 M_{\Sol}$.
Using this estimate, binary masses of $3.3$ and $2.5 
M_{\Sol}$ \citep{hil72},
and the 2.2~yr period, the semimajor axis of the orbit would be
3.2~AU. This corresponds to a light-travel time of 27 minutes.
Combining this with the amplitude of 4.18~minutes measured
by \citet{koz99a} yields an orbital inclination of $9\degr$ with 
respect to the line of sight, so that nearly the full 3.2 AU is 
visible to the observer. Assuming a distance of $480\,{\rm pc}$,
the maximum angular elongation is $0\arcsec.007$. This is twice the 
resolution limit of the Keck interferometer operating at
$1.5\,{\rm \mu m}$
with its $85\,{\rm m}$ baseline. The interpretation of eclipse timings 
by \citet{koz99a} and \citet{koz99b} 
also predicts the times of maximum elongation.

In the case of DI Herculis, no indirect indication of third light 
exists. However, information from past theoretical analyses provides hope that present 
interferometers should be capable of directly observing the 
putative third-body. The analysis of \citet{kha91} suggests that the 
minimum third-body mass is about $0.8 M_{\Sol}$ and that the minimum period is 
about 7~yr. Assuming binary masses of $5.15$ and $4.52 
M_{\Sol}$ 
\citep{pop82}, we conclude that the semimajor axis is at least 8 AU. 
Since the orbit is expected to be highly inclined, and the distance 
to DI Her is about $500\,{\rm pc}$, we expect a maximum angular elongation of
$0\arcsec.02$.

Of course, the ability to resolve these third bodies is of little use 
unless they are sufficiently bright. Here infrared observations are a 
great advantage, since the compact binary stars are relatively massive 
in comparison with the hypothesized third bodies. For example, 
applying the 
mass-luminosity relationship to AS Cam, we conclude that the third 
star should have a total luminosity less than the system by 
4.3~mag. 
A simple calculation based on the Planck distribution and assumed 
temperatures of $20,000$ and $6000\,{\rm K}$ for the binary and third 
star, respectively, predicts that, in the $H$ ($1645\pm 155\,{\rm nm}$) 
and $K$ 
($2200\pm 480\,{\rm nm}$) bands, the third star 
is dimmer by only about 1~mag. A similar calculation for DI 
Her predicts that in the $H$ and $K$ bands the third star is dimmer than the 
system by 3.5~mag. Finally, the magnitudes of these systems are such that the 
compact binary stars may serve as natural guide stars for adaptive 
optics in the case of the Keck interferometer.

Bolstered by indirect evidence in the case of AS Cam and dynamical 
stability indicated by the considerations of this paper,
the case for a third-body solution to the long standing problem of 
anomalous apsidal motion is stronger than ever. The final judgment, 
however, may soon be expected from the current generation of 
interferometers.

\clearpage

\clearpage

\clearpage

\clearpage

\begin{table}
%\begin{center}
\caption{Stability criteria for AS Cam and DI Her. \label{Table1}}
\begin{tabular}{ccccc}
\tableline\tableline
System & $S_{tb}^{*}$ & $S_{cl}$ & $S_{tb}^{**}$ & $S_{rel}$ \\
\tableline
AS Cam &$3\cdot 10^{-6}$ & $4.4\cdot 10^{-6}$ & $3.8\cdot 10^{-8}$ & 
$10^{-8}$ \\
DI Her & $4\cdot 10^{-7}$ & $10^{-7}$ & $(0.7-1.5)\cdot 10^{-8}$ & $10^{-8}$ \\
\tableline
\end{tabular}
%\end{center}
\end{table}

\end{document}